\documentstyle[aps,prb,epsf,multicol]{revtex}

\newif\ifmynarrow \mynarrowfalse

\renewcommand{\narrowtext}{%
  \ifmynarrow\hspace*{\fill}\raisebox{-1ex}[0pt][0pt]{%
    \rule{0.3pt}{1ex}%
    \rule[1ex]{20.5pc}{0.3pt}}\fi
  \mynarrowtrue
  \vspace{-1.0ex}%
  \begin{multicols}{2}%
  \par\global\columnwidth20.5pc
  \global\hsize\columnwidth\global\linewidth\columnwidth
  \global\displaywidth\columnwidth}

\renewcommand{\widetext}{%
  \end{multicols}%
  \vspace{-2.5ex}%
  \noindent\raisebox{1ex}[0pt][0pt]{%
    \rule{20.5pc}{0.3pt}%
    \rule{0.3pt}{1ex}}%
  \par\global\columnwidth42.5pc
  \global\hsize\columnwidth\global\linewidth\columnwidth
  \global\displaywidth\columnwidth}

\newcommand{\ak}{{\bf k}}
\newcommand{\akb}{k}
\newcommand{\ac}{c}
\newcommand{\abasis}{\chi}
\newcommand{\aansatz}{\phi}
\newcommand{\aBZ}{\text{BZ$_{0}$}}
\newcommand{\pnn}{{\bf p}^{{\bf k}_{0}}_{n n'}}

\newcommand{\nk}{{\bf K}}
\newcommand{\nkb}{K}
\newcommand{\nc}{C}
\newcommand{\nbasis}{{\rm X}}
\newcommand{\nansatz}{\Phi}
\newcommand{\nBZ}{\text{BZ$_{1}$}}
\newcommand{\PnnK}{{\bf P}^{\set}_{{\it n n'}}}
\newcommand{\pnnK}{{\bf p}^{\set}_{n n'}}

\newcommand{\set}{\bbox {\cal K}}
\newcommand{\VnnKK}{V^{\set \set'}_{n n'}}

\newcommand{\rr}{{\bf r}}
\newcommand{\kdotp}{{\bf k \cdot p}}
\newcommand{\PG}{P^{\Gamma}}
\newcommand{\PL}{P^{\text{L}}}
\newcommand{\ex}{{\bf \protect\hat e}_{x}}
\newcommand{\ey}{{\bf \protect\hat e}_{y}}
\newcommand{\ez}{{\bf \protect\hat e}_{z}}
\newcommand{\signum}{{\rm sgn}}

\begin{document}
\draft

\title{Generalization of $\kdotp$ theory for periodic perturbations}
\author{R.~Stubner,\cite{rs} R.~Winkler, and O.~Pankratov}
\address{Institut f\"ur Technische Physik III, Universit\"at
  Erlangen-N\"urnberg, \\
Staudtstr.~7, D-91058 Erlangen, Germany}
\date{\today}
\maketitle

\begin{abstract}
  We extend standard $\kdotp$ theory to take into account periodic
  perturbations which are rapidly oscillating with a wavelength of a few
  lattice constants. Our general formalism allows us to explicitly
  consider the Bragg reflections due to the perturbation-induced
  periodicity.  As an example we calculate the effective masses in the
  lowest two conduction bands of spontaneously ordered GaInP$_{2}$ as a
  function of the degree of ordering. Comparison of our results for the
  lowest conduction band to available experimental data and to first
  principle calculations shows good agreement.
\end{abstract}

\narrowtext

\section{Introduction}
\label{sec:intro}

For many years $\kdotp$ theory\cite{kane:66,call:74} has been very
successful in describing a wide variety of crystal band structures, an
important merit being that it allows one to derive simple, analytical
formulas which capture the essential physics. In
the presence of perturbing potentials it becomes the envelope function
approximation (EFA),\cite{luko:55} which has been successfully applied to
such different problems as impurities\cite{kohn:57} and semiconductor
heterostructures.\cite{bast:88} However, the EFA requires the
perturbation potential $V_{1}(\rr)$ to be slowly varying on the length
scale of the lattice constant, i.e., the nonzero Fourier components of
the potential $V_{1}(\rr)$ must be restricted to wave vectors which are
small compared to the dimensions of the Brillouin zone. Some
authors\cite{wozu:96,fwz:95} reported problems with the EFA for systems
where this requirement is not fulfilled, notably in artificial and
natural short-period superlattices such as spontaneously ordered
GaInP$_{2}$.  These superlattices can be viewed as systems with a
periodic perturbation, where the smallest Fourier components of the
perturbation are comparable to the dimensions of the Brillouin zone of
the unperturbed problem. These Fourier components result in interactions
between states with very different wave vectors, which are crucial for
the properties of these short-period superlattices. However, the
interactions cannot be accurately described within standard $\kdotp$
theory.\cite{luko:55} In this paper we present a general method which
allows us to treat perturbing potentials which are rapidly oscillating
but commensurate to the periodicity of the potential of the unperturbed
problem.  To illustrate the method, we apply it to the natural
superlattice of spontaneously ordered GaInP$_{2}$.

The Ga$_{x}$In$_{1-x}$P alloy for $x \approx 0.51$ can be
lattice matched grown on a GaAs$(001)$ substrate. Under proper growth
conditions, long-range order of the CuPt type is
observed.\cite{gsi:88,zuma:94} This type of ordering is characterized by
$\langle 111 \rangle$ layers alternatingly rich in Ga or In. In
the ordered material the symmetry of the lattice is reduced from $T_{d}$
to $C_{3v}$, and the Brillouin zone becomes smaller than the zinc-blende
Brillouin zone, which leads to a backfolding of states. The change in the
crystal potential induced through the ordering is a short-period
potential that mixes electronic states of an ``averaged'' zinc-blende
structure.  In particular, the interactions between $\Gamma$ and L
states lead to energy shifts of the band-edge states in the ordered
alloy, which cause band-gap reduction and valence-band splitting. These
effects have been investigated theoretically\cite{wezu:89,wezu:98} and
experimentally.\cite{gsi:88,fzcm:97,fgmz:98} In addition to the changes
in the energies, the effective masses are also altered. Raikh and
Tsiper\cite{rats:94} calculated the conduction-band effective mass of
ordered GaInP$_{2}$ using a two-band model, which accounts only for the
mixing of conduction-band $\Gamma$ and L states. They found that the
effective mass parallel to the ordering direction $m_{\parallel}$ and
the effective mass perpendicular to the ordering direction $m_{\perp}$
increase with increasing ordering, and that $m_{\parallel}$ is larger
than $m_{\perp}$. This model, however, does not take into account the
change of the interaction between conduction- and valence-band states due
to the band-gap reduction and valence-band splitting. These changes were
investigated by Zhang and Mascarenhas\cite{zhma:95} with an eight-band
$\kdotp$ model, which included zinc-blende $\Gamma$ states from both the
conduction and valence band. They find $m_{\parallel}$ and
$m_{\perp}$ to decrease with increasing ordering, with $m_{\parallel}$
being larger than $m_{\perp}$. A third investigation was done by
Franceschetti, Wei, and Zunger,\cite{fwz:95} who performed
first-principle calculations using the local-density approximation. They
find 
$m_{\parallel}$ to increase, whereas $m_{\perp}$ decreases with
increasing ordering.  This result agrees qualitatively with the only
measurment that investigated the anisotropy of the effective masses in
ordered GaInP$_{2}$.\cite{ezdm:97} In Ref.~\onlinecite{fwz:95} the
conclusion was drawn that the conduction-band effective masses in ordered
GaInP$_{2}$ depend on a ``delicate balance'' of $\Gamma$--L mixing and
increased interaction between conduction and valence band. However, the
$\Gamma$--L mixing and the increase of the interaction between
conduction and valence bands have a common source in the ordering-induced
$\Gamma$--L interactions.  The conduction band masses in partially
ordered GaInP$_{2}$ present an excellent test for our theory, which
should be able to describe that ``delicate balance.''

This paper is organized as follows. In Sec.~\ref{sec:gentheo} we derive
a general scheme applicable to periodic perturbations within $\kdotp$
theory. In Sec.~\ref{sec:application} we use this theory to derive a
model for the conduction band of spontaneously ordered GaInP$_{2}$. The
results are discussed in Sec.~\ref{sec:results}. A summary and a short
outlook are presented in Sec.~\ref{sec:summary}.

\section{General Theory}
\label{sec:gentheo}

We consider a system with the one-particle Hamiltonian
\begin{equation}
  \label{eq:H0}
  H_{0} = \frac{{\bf p}^{2}}{2m} + V_{0}({\rr}),
\end{equation}
where ${\bf p}$ is the momentum operator, $m$ the free-electron mass, and
$V_{0}({\rr})$ the periodic potential of the crystal. In the case of the
GaInP$_{2}$ alloy, $V_{0}({\rr})$ is an averaged potential of the
disordered material. The eigenfunctions of the Hamiltonian $H_{0}$ are
Bloch functions $\psi_{n \ak}({\rr}) = e^{i \ak \cdot \rr} u_{n
  \ak}({\rr})$ with eigenvalues $\varepsilon_{n}(\ak)$
\begin{equation}
  \label{eq:SG-H0}
  H_{0} \psi_{n \ak} = \varepsilon_{n}(\ak) \psi_{n \ak}.
\end{equation}
Here $\ak$ is a wave vector in the first Brillouin zone ($\aBZ$), which
corresponds to the periodicity of $V_{0}({\rr})$.

In conventional $\kdotp$ theory, basis functions
\begin{equation}
  \label{eq:abasis}
  \abasis^{\ak_{0}}_{n \ak}(\rr) = e^{i \ak \cdot \rr} \psi_{n \ak_{0}}(\rr) 
  = e^{i \ak \cdot \rr} e^{i \ak_{0} \cdot \rr} u_{n \ak_{0}}(\rr)
  ,
\end{equation}
are introduced, where $\ak$ belongs to $\aBZ$.  As shown by Luttinger
and Kohn,\cite{luko:55} the functions $\abasis^{\ak_{0}}_{n \ak}$
provide a complete and orthonormal basis set for any $\ak_{0}$ being
an element of $\aBZ$. Thus any eigenfunction of $H_{0}$ can be expanded in
terms of $\abasis^{\ak_{0}}_{n \ak}$
\begin{equation}
  \label{eq:aansatz}
  \aansatz_{n''}(\rr) = \sum_{n'} \int\limits_{\aBZ} d^{3} \akb' \, 
  \ac_{n'' n'}(\ak') \abasis^{\ak_{0}}_{n' \ak'}({\rr}) 
  ,
\end{equation}
which yields the well-known $\kdotp$ equation for expansion coefficients
$\ac_{n'' n}(\ak)$ and energy $\varepsilon_{n''}(\ak_{0} + \ak)$
\begin{equation}
  \label{eq:SG-standard}
  \sum_{n'} H^{\ak_{0}}_{n n'}(\ak) \ac^{}_{n'' n'}(\ak) = 
  \varepsilon_{n''}(\ak_{0} + \ak) \ac_{n'' n}(\ak) ,
\end{equation}
with
\begin{equation}
  \label{eq:Hnn}
  H^{\ak_{0}}_{n n'}(\ak) = \left( \varepsilon_{n}(\ak_{0}) +
    \frac{\hbar^{2}}{2m} \akb^{2} \right) \delta_{n n'} +
    \frac{\hbar}{m} \ak \cdot \pnn.
\end{equation}
The momentum matrix element $\pnn$ is defined as
\begin{equation}
  \label{eq:pnn}
  \pnn = {\frac{(2\pi)^{3}}{\Omega_{0}}} \int d^{3}r \,
  e^{-i \ak_{0} {\bf \cdot r}} u_{n\ak_{0}}^{\ast}  
  {\bf p} e^{i \ak_{0} {\bf \cdot r}} u_{n'\ak_{0}}.
\end{equation}
Here the integration extends over the unit cell with volume
$\Omega_{0}$.  Equation~(\ref{eq:SG-standard}) is diagonal with respect
to $\ak$ due to the periodicity of $V_{0}(\rr)$, i.e., $\ak$ is a good
quantum number and the eigenstates $\aansatz_{n}$ can be written as a
superposition of basis functions $\abasis^{\ak_{0}}_{n \ak}$ for
different bands $n$ but the same wave vector $\ak$.  Because of the
completeness of the basis functions $\abasis^{\ak_{0}}_{n \ak}(\rr)$ an
arbitrary potential $V_{1}({\rr})$ can be taken into account in
Eq.~(\ref{eq:SG-standard}), even if $V_{1}$ does not have the
periodicity of $V_{0}$. However, this results in eigenstates
$\aansatz_{n}(\rr)$ which are superpositions of functions
$\abasis^{\ak_{0}}_{n \ak}$ for different $\ak$. In practical
calculations one is usually restricted to wave vectors $\ak+\ak_{0}$
close to $\ak_{0}$, since the $\kdotp$ term is treated as a
perturbation.  Therefore, this method will fail if the additional
potential $V_{1}({\rr})$ is not smooth.

Here we will generalize the above approach to perturbing potentials
$V_{1}({\rr})$ which are rapidly oscillating but periodic and
commensurate to the periodicity of $V_{0}$. In the case of GaInP$_{2}$,
$V_{1}({\rr})$ corresponds to the ordering potential.\cite{wezu:89}

The total potential $V_{0} + V_{1}$ is characterized by a larger unit
cell, and hence a smaller Brillouin zone ($\nBZ$). Due to commensurability,
the reciprocal-lattice vectors associated with the original potential
$V_{0}$ can be expressed as an integer linear combination of the
reciprocal-lattice vectors associated with the perturbing potential. In
particular we have {\em several\/} wave vectors in the larger Brillouin
zone $\aBZ$ which become reciprocal-lattice vectors of the perturbed
problem, i.e., these wave vectors become equivalent to the $\Gamma$
point. This set of wave vectors will be called $\{\set\}$. For ordered
GaInP$_{2}$ this consists of the $\Gamma$ and L point of $\aBZ$.

The main difference from standard $\kdotp$ theory is that eigenfunctions
of $H_{0}$ belonging to wave vectors of the {\em set\/} $\{\set\}$ are
used to form basis functions of the form
\begin{equation}
  \label{eq:nbasis}
  \nbasis_{n \nk}^{\set}(\rr) = e^{i \nk \cdot \rr} \psi_{n \set}(\rr) 
  = e^{i \nk \cdot \rr} e^{i \set \cdot \rr} u_{n \set}({\rr}).
\end{equation}
Any function having the periodicity of the perturbed system can be
expanded in terms of the functions $\psi_{n \set}(\rr)$ as the set
$\{\set\}$ is folded onto the $\Gamma$ point of $\nBZ$. Therefore, when
$\nk$ is taken from $\nBZ$ the functions $\nbasis_{n \nk}^{\set}(\rr)$
form a {\em complete and orthonormal\/} basis  (cf.\ also
Refs.~\onlinecite{luko:55} and~\onlinecite{fore:98}). The eigenfunctions
of $H = H_{0} + V_{1}$ are expanded in terms of~$\nbasis_{n
  \nk}^{\set}$,
\begin{equation}
  \label{eq:nansatz}
  \nansatz_{n''}(\rr) = \sum_{\set',n'} \int\limits_{\nBZ} d^{3} \nkb' \,
  \nc_{n'' n'}^{\set'}(\nk') \nbasis_{n' \nk'}^{\set'}(\rr).
\end{equation}
This yields the Schr\"odinger equation for the expansion
coefficients $\nc_{n'' n}^{\set}(\nk)$,
\widetext
\begin{equation}
  \label{eq:SG}
  \sum_{n'} \left[ \left( \varepsilon_{n}(\set) + \frac{\hbar^{2}}{2m}
  \nkb^{2} \right) \delta_{n n'} + \frac{\hbar}{m} {\bf \nk \cdot \PnnK}
  \right] \nc^{\set}_{n'' n'} (\nk) + \sum_{n',\set'}  \VnnKK \nc^{\set'}_{n''
  n'} (\nk)  =  E^{}_{n''}(\nk) \nc_{n'' n}^{\set}(\nk) ,
\end{equation}
where
\begin{equation}
  \label{eq:PnnK}
  \PnnK = {\frac{(2\pi)^{3}}{\Omega_{1}}} \int d^{3}r \, e^{-i{\bf \set \cdot
  r}} u_{n\set}^{\ast}  {\bf p} e^{i{\bf \set \cdot r}} u_{n'\set}  
\end{equation}
and
\begin{equation}
  \label{eq:VnnKK}
  \VnnKK =  {\frac{(2\pi)^{3}}{\Omega_{1}}} \int d^{3}r \,
  e^{-i {\bf \set \cdot r}} u_{n\set}^{\ast} V_{1}  e^{i{\bf \set' \cdot
  r}} u_{n'\set'}.
\end{equation}
On the left-hand side of Eq.~(\ref{eq:SG}) the part diagonal in $\set$
is identical to $H_{n n'}^{\set}$ defined in Eq.~(\ref{eq:Hnn}), because
we have $\PnnK = \pnnK$. [Note that in Eq.~(\ref{eq:PnnK}) the larger
normalizing volume $\Omega_{1}$ is compensated for by the larger integration
volume.] The only effect of the perturbing potential $V_{1}$ are the
coupling matrix elements $\VnnKK$, and Eq.~(\ref{eq:SG}) can be written
as
\begin{equation}
  \label{eq:SG-Hnn}
  \sum_{n'} H^{\set}_{n n'}(\nk) \nc^{\set}_{n'' n'}(\nk) + \sum_{n' \set'}
  \VnnKK \nc^{\set'}_{n'' n'} (\nk)  
  =  E^{}_{n''}(\nk) \nc_{n'' n}^{\set}(\nk). 
\end{equation}
\narrowtext
\noindent 
Equation~(\ref{eq:SG}) represents our generalization of the standard
$\kdotp$ equation (\ref{eq:SG-standard}).  Like
Eq.~(\ref{eq:SG-standard}), this new equation depends only {\em
  explicitly\/} on $\nk$.\cite{notation}

As noted above, the set $\{\set\}$ is folded onto the $\Gamma$ point of
$\aBZ$.  Therefore, an expansion in terms of the basis
functions~(\ref{eq:nbasis}) is appropriate for the description of states
near the center of the Brillouin zone of the perturbed system. If one is
interested in states near $\nk_{0} \ne 0$, one has to replace $\set$ by
$\set + \nk_{0}$ in Eq.~(\ref{eq:nbasis}).

It is straightforward to include spin-orbit coupling and strain-induced
effects\cite{bipi:74} in Eq.~(\ref{eq:SG}). Likewise, we can easily add
a slowly varying perturbation, like the potential of an impurity, as
ansatz (\ref{eq:nansatz}) does not require periodicity.

\section{Application to partially ordered G\lowercase{a}I\lowercase{n}P$_{2}$}
\label{sec:application}

In this section we apply the above theory to the conduction-band
effective masses at the $\Gamma$ point of partially ordered GaInP$_{2}$.
The unperturbed potential $V_{0}$ corresponds to the disordered
material.  The perturbation $V_{1}$ represents the ordering potential
which is defined in Ref.~\onlinecite{wezu:89} as the difference between
the potentials of the ordered and disordered material. In principle
there are four equivalent variants of CuPt ordering for GaInP$_{2}$,
corresponding to the four $\langle 111 \rangle$ directions. Due to
substrate effects only two of them are observed in experiments,
however.\cite{gsi:88} Since we consider the bulk system, the domain
structure is not relevant to our calculations. Hence we choose the
$[111]$ direction to be parallel to the ordering direction.

Due to the periodicity of the ordering potential $V_{1}$, zone center
states of ordered GaInP$_{2}$ are derived from zinc-blende $\Gamma$-  and
L-point states of the disordered system. Thus we choose \{$\Gamma$, L\}
as the wave-vector set $\{\set\}$. We restrict ourselves to a seven-band
model, containing the zinc-blende $\Gamma_{1 \text{c}}$, $\Gamma_{5
  \text{v}}$, L$_{1 \text{c}}$, and L$_{3 \text{v}}$ states
(nomenclature according to Koster {\em et al}.\cite{kdws:63}).
Spin-orbit interaction is neglected, as it has only a minor influence
on the conduction-band effective masses.
Due to
time-reversal symmetry wave functions from $\Gamma$ {\em and\/} L points
can be chosen to be real. With this phase convention all momentum and
potential matrix elements can be defined as real quantities.

\subsection{$\kdotp$ Hamiltonian}
\label{sec:k.p-H}

The Hamiltonian $H^{\Gamma}$ describing the $\kdotp$ interaction between
$\Gamma_{5 \text{v}}$ and $\Gamma_{1 \text{c}}$ is well
known.\cite{kane:66} We are only interested in conduction-band effective
masses, so we do not consider remote band contributions in the valence
band. In the conduction band, both the $\kdotp$ interaction with the
topmost valence band and remote bands contribute to the effective mass.
However, the latter terms are rather small, and we neglect them here.
Hence $H^{\Gamma}$ reads as follows
\begin{equation}
  \label{eq:HG}
  H^{\Gamma}(\nk) = \left(
\begin{array}{cccc}
 E^{\Gamma}_{\text{c}} + \frac{\hbar^{2}}{2m} \nkb^{2} 
&  i\PG \nkb_{x} 
&  i\PG \nkb_{y} 
&  i\PG \nkb_{z} \\
  -i\PG \nkb_{x} 
& \frac{\hbar^{2}}{2m} \nkb^{2}
& 0
& 0 \\
  -i\PG \nkb_{y} 
& 0
& \frac{\hbar^{2}}{2m} \nkb^{2} 
& 0 \\
  -i\PG \nkb_{z} 
& 0
& 0
& \frac{\hbar^{2}}{2m} \nkb^{2} \\
\end{array}\right).
\end{equation}
The energy reference in Eq.~(\ref{eq:HG}) is taken at the maximum of the
valence band.  Note that $H^{\Gamma}$ is spherically symmetric, i.e., we
can choose the coordinate system to be $\ex \parallel [1 \bar{1} 0]$,
$\ey \parallel [1 1 \bar{2}]$ and $\ez \parallel [111]$, which is
convenient for describing the L point. The only parameters we need to
know to specify Eq.~(\ref{eq:HG}) are the band gap
$E^{\Gamma}_{\text{c}}$ and Kane's momentum matrix element\cite{kane:66}
\begin{equation}
  \label{eq:PG}
    \PG = - i \frac{\hbar}{m} \langle \Gamma_{1 \text{c}} | p_{x}
    | \Gamma_{5 \text{v}}^{x} \rangle.
\end{equation}

The situation at the L point is very similar, in that there is
only one reduced matrix element 
\begin{equation}
  \label{eq:PL}
  \PL = - i \frac{\hbar}{m} \langle \text{L}_{1 \text{c}} | p_{x}
    | \text{L}_{3 \text{v}}^{x} \rangle.
\end{equation}
If we neglect remote band contributions in the conduction band, $\PL$ is
responsible for the conduction-band transverse mass
$m^{\text{L}}_{\perp}$.  However, interactions between $\text{L}_{1
  \text{c}}$ and $\text{L}_{3 \text{v}}$ states cannot account for the
conduction-band longitudinal mass $m^{\text{L}}_{\parallel}$, and the
longitudinal mass would be equal to the free-electron mass $m$ without
contributions from remote bands. Therefore, in the L point $\kdotp$
matrix we have to retain the parameter $G$, which represents remote band
contributions to $m^{\text{L}}_{\parallel}$.  Neglecting remote band
contributions in the valence band, the Hamiltonian matrix $H^{\text{L}}$
takes the form
\widetext
\begin{equation}
  \label{eq:HL}
  H^{\text{L}}(\nk) = \left(\begin{array}{ccc}
E^{\text{L}}_{\text{c}} +\frac{\hbar^{2}}{2m} \nkb^{2} + G \nkb_{z}^{2}
& i\PL \nkb_{x} 
& i\PL \nkb_{y} \\
  -i\PL \nkb_{x} 
& E^{\text{L}}_{\text{v}} + \frac{\hbar^{2}}{2m} \nkb^{2} 
& 0 \\ 
  -i \PL \nkb_{y} 
& 0 
& E^{\text{L}}_{\text{v}} + \frac{\hbar^{2}}{2m} \nkb^{2} \\ 
\end{array}\right). 
\end{equation}
\narrowtext

\subsection{Matrix elements of $V_{1}$}
\label{sec:V-me}

The potential of the ordered material can be modeled by dividing the
lattice into two sublattices which are rich in Ga or In, respectively,
and averaging separately over the two sublattices. Subtracting the
averaged potential of the disordered material, one obtains a model for the
ordering potential as used in Ref.~\onlinecite{rats:94}.  In principle,
there are two different types of matrix elements of the ordering
potential, those which couple $\Gamma$- and L-point states, and those
which lead to interactions within $\Gamma$- or L-point states,
respectively. However, the latter matrix elements are exactly zero, if
the ordering potential is modeled with the above outlined separate
virtual crystal approximations over the two sublattices.  Therefore, it
can be expected that these matrix elements are small, and they are
neglected here. The nonzero matrix elements of the ordering potential
$V_{1}$ can be derived using group theory, and we are left with only
three real reduced matrix elements
\begin{mathletters}
  \label{eq:def-V}
  \begin{eqnarray}
    \label{eq:V11,V35,V15}
    V_{11} &=& \langle \Gamma_{1\text{c}} | V_{1} |
    \text{L}_{1\text{c}} \rangle\\  
    V_{35} &=& \langle \Gamma_{5\text{v}}^{x} | V_{1} |
    \text{L}_{3\text{v}}^{x} \rangle = \langle \Gamma_{5\text{v}}^{y} |
    V_{1} | \text{L}_{3\text{v}}^{y} \rangle\\
    V_{15} &=& \langle \Gamma_{5\text{v}}^{z} | V_{1} |
    \text{L}_{1\text{c}} \rangle.
  \end{eqnarray}
\end{mathletters}
These equations illustrate that our generalized approach shares the
well-known and important feature of standard $\kdotp$ theory that by
means of group theory the number of independent parameters can be
greatly reduced.

\begin{figure}[h]
\epsfxsize=0.3\textwidth \epsfbox{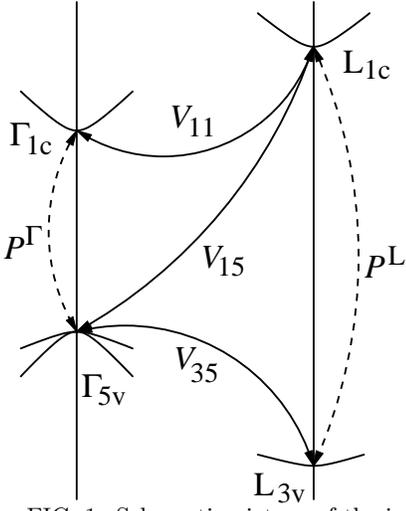}
  \caption{Schematic picture of the interactions described by the
    momentum and potential matrix elements in Eq.~(\ref{eq:HGL}).}
  \label{fig:P-und-V}
\end{figure}

Combining Eqs.~(\ref{eq:HG}), (\ref{eq:HL}), and (\ref{eq:def-V})  we
end up with a Hamiltonian of the form 
\begin{equation}
  \label{eq:HGL}
  H^{\Gamma \text{L}}(\nk) = \left(
  \begin{array}{cc}
H^{\Gamma}(\nk) & V^{\Gamma \text{L}} \\
{V^{\Gamma \text{L}}}^{\dagger} & H^{\text{L}}(\nk)
  \end{array}
\right)
\end{equation}
with
\begin{equation}
  \label{eq:VGL}
  V^{\Gamma \text{L}} = \left(
    \begin{array}{ccc}
      V_{11} & 0 & 0 \\
      0 & V_{35} & 0 \\
      0 & 0 & V_{35} \\
      V_{15} & 0 & 0 
    \end{array} \right).
\end{equation}
Figure~\ref{fig:P-und-V} shows a schematic picture of the interactions
described by the different matrix elements. 

\subsection{Values of the matrix elements}
\label{sec:values}

Two limiting cases are used to determine the numerical values of the
potential and momentum matrix elements in Eq.~(\ref{eq:HGL}). For
$V_{11} = V_{15} = V_{35} = 0$ the model describes the disordered
material, and the unknown parameters $\PG$, $\PL$ and $G$ can be fitted
to the conduction-band effective masses at the $\Gamma$ and L points,
respectively. Such an analysis using experimental data has been done for
the $\Gamma$ point,\cite{edhm:94} but not for the L point. In order to
obtain a consistent set of parameters, we deduce the effective masses and
band gaps from a band-structure calculation based on an empirical
tight-binding model with $sp^{3}d^{5}s^{\ast}$ basis, nearest neighbor
interactions, and without spin-orbit interaction.  Jancu {\em et
  al.}\cite{jsbb:98} showed that a tight-binding model with such a basis
is capable of accurately describing the valence bands and the two lowest
conduction bands in many diamond and zinc-blende-type semiconductors.
The tight-binding parameters we use are interpolated from the values for
GaP and InP in Ref.~\onlinecite{jsbb:98}, with a Ga:In ratio of 51:49. In
order to correctly reproduce the fundamental band gap in this
virtual-crystal approximation, we incorporate an empirical bowing factor
$b_{(ll'm)} = 1/2 [(ll'm)_{\text{GaP}} - (ll'm)_{\text{InP}} ]$ for the
four $(ss\sigma)$-type two-center integrals. Band gaps and effective
masses from this calculation and the resulting values for $\PG$, $\PL$
and $G$ are summarized in Table~\ref{tab:E-m-P}. We use a phase
convention for the wave functions, such that both $\PG$ and $\PL$ are
positive.  The values for $\PG$ and $\PL$ are very close to each
other,\cite{card:63} so we use the approximation
\begin{equation}
  \label{eq:PG-gleich-PL}
  \PL = \PG = 8.86\text{~eV\AA}.
\end{equation}

With nonzero potential matrix elements but $\nk = 0$, the Hamiltonian
matrix (\ref{eq:HGL}) describes the zone center states of ordered
GaInP$_{2}$. These states have been studied previously, both
experimentally\cite{fzcm:97,fgmz:98,priv,kksk:99} and
theoretically.\cite{wfz:95,wezu:98} These studies indicate that there is
a certain correlation between different ordering-induced changes of the
band structure.  In particular, the crystal-field splitting
$\Delta_{\text{CF}}$, the band-gap reduction $\Delta E_{\text{BGR}}$,
and the change in the transition energy $\Delta E_{\Gamma \rightarrow
  \text{L}}$ for the ordering-induced transition
$\bar\Gamma_{3\text{v}}(\Gamma_{5\text{v}}) \rightarrow
\bar\Gamma_{1\text{c}}(\text{L}_{1\text{c}})$ have a fixed ratio for all
samples:\cite{theta}
\begin{mathletters}
  \label{eq:ratio}
  \begin{eqnarray}
    \label{eq:zeta}
    \zeta &=& \frac{\Delta E_{\text{BGR}}}{\Delta_{\text{CF}}} = 2.66,\\
    \label{eq:theta}
    \theta &=& \frac{\Delta E_{\Gamma \rightarrow \text{L}}}{\Delta
    E_{\text{BGR}}} = 0.48.
  \end{eqnarray}
\end{mathletters}
Expressing $\Delta_{\text{CF}}$, $\Delta E_{\text{BGR}}$, and $\Delta
E_{\Gamma \rightarrow \text{L}}$ as functions of $V_{11}$, $V_{15}$,
and $V_{35}$ to second order in these matrix elements and using
ratios~(\ref{eq:ratio}), we obtain
\begin{mathletters}
\label{eq:V-V}
\begin{eqnarray}
  \label{eq:V35-V11}
  \frac{|V_{35}|^{2}}{E^{\text{L}}_{\text{v}}} &=&
  - 0.426 \frac{|V_{11}|^{2}}{E^{\text{L}}_{\text{c}}-E^{\Gamma}_{\text{c}}} \\
  \label{eq:V15-V11}
  \frac{|V_{15}|^{2}}{E^{\text{L}}_{\text{c}}} &=&  + 0.110
  \frac{|V_{11}|^{2}}{E^{\text{L}}_{\text{c}}-E^{\Gamma}_{\text{c}}}%
.
\end{eqnarray}
\end{mathletters}

Equation~(\ref{eq:V-V}) determines the relation between $|V_{11}|$ and
$|V_{35}|$, and between $|V_{11}|$ and $|V_{15}|$. Different degrees of
ordering, i.e., different strengths of the ordering potential, can
therefore be modeled by different values of $V_{11}$. The matrix element
$V_{11}$ itself is proportional to the degree of ordering $\eta$, as
defined in Ref.~\onlinecite{rats:94}, if the ordering potential
$V_{1}(\rr)$ is described by separate virtual-crystal
approximations over two sublattices described above.  Note that this
method does not determine the relative signs of the matrix
elements in (\ref{eq:def-V}).

\subsection{Diagonalization}
\label{sec:diag}

The band-gap reduction in highly ordered samples is about
150~meV.\cite{ezdm:97,fzcm:97,fgmz:98} This corresponds to $V_{11}
\approx 200$~meV in our model.  Thus, according to
Eq.~(\ref{eq:V15-V11}), the potential matrix element $V_{15}$, which
couples L$_{1 \text{c}}$ and $\Gamma_{5 \text{v}}^{z}$ states, is small
compared to the energy difference between these states. We therefore use
L\"owdin perturbation theory\cite{lowd:51} to calculate the change in
energy of these states to second order. For the L$_{1 \text{c}}$ state
this gives $E^{\text{L}}_{\text{c}} +
|V_{15}|^{2}/E^{\text{L}}_{\text{c}}$, whereas for the $\Gamma_{5
  \text{v}}^{z}$ state the energetic position of the level is
$\tilde{E}^{\Gamma z}_{\text{v}} = -
|V_{15}|^{2}/E^{\text{L}}_{\text{c}}$.  Neglecting the mixing of wave
functions, we decouple valence and conduction band with respect to the
ordering potential by this procedure.

\begin{figure}[h]
\epsfxsize=0.45\textwidth \epsfbox{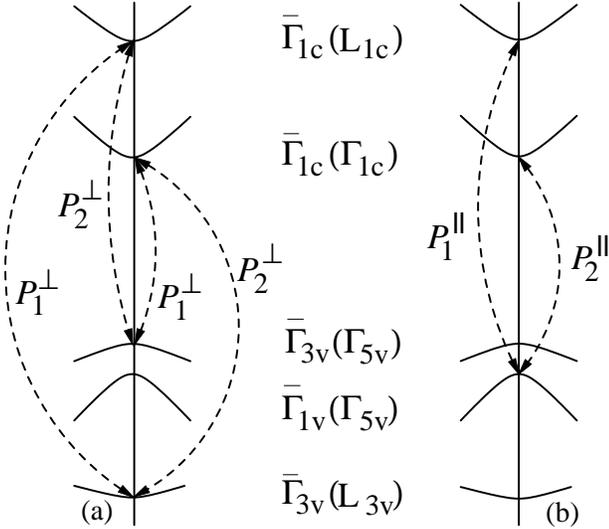}
  \caption{Schematic picture of the interactions described by the
    momentum matrix elements after diagonalization with respect to the
    ordering potential [Eq.~(\ref{eq:HGbar})]. For clarity the picture
    is split into (a) $\nk$ perpendicular and (b) $\nk$ parallel to the
    ordering direction.} 
  \label{fig:P-und-P}
\end{figure}

The problem thus reduces to two two-level systems, which can be solved
analytically, resulting in energy eigenvalues $E^{(1/2)}_{\text{c}}$ and
$E^{(1/2)}_{\text{v}}$, and expansion coefficients for the zone-center
states in the conduction band,
\begin{mathletters}
  \label{eq:CB_expansion}
  \begin{eqnarray}
    \label{eq:CB_expansion_G}
    | \bar\Gamma_{1 \text{c}}(\Gamma_{1 \text{c}}) \rangle &=&
      \alpha_{\text{c}} |\Gamma_{1 \text{c}} \rangle +
      \beta_{\text{c}} |\text{L}_{1 \text{c}} \rangle,\\
    \label{eq:CB_expansion_L}
    | \bar\Gamma_{1 \text{c}}(\text{L}_{1 \text{c}}) \rangle &=&
    \beta_{\text{c}} | \Gamma_{1 \text{c}} \rangle - 
    \alpha_{\text{c}}| \text{L}_{1 \text{c}} \rangle,
  \end{eqnarray}
\end{mathletters}
and in the valence band,
\begin{mathletters}
  \label{eq:VB_expansion}
  \begin{eqnarray}
    \label{eq:VB_expansion_G}
    | \bar\Gamma^{x}_{3 \text{v}}(\Gamma^{x}_{5 \text{v}}) \rangle &=&
      \alpha_{\text{v}} |\Gamma^{x}_{5 \text{v}} \rangle +
      \beta_{\text{v}} |\text{L}^{x}_{3 \text{v}} \rangle,\\
    \label{eq:VB_expansion_L}
    | \bar\Gamma^{x}_{3 \text{v}}(\text{L}^{x}_{3 \text{v}}) \rangle &=&
    \beta_{\text{v}} | \Gamma^{x}_{5 \text{v}} \rangle - 
    \alpha_{\text{v}}| \text{L}^{x}_{5 \text{v}} \rangle.
  \end{eqnarray}
\end{mathletters}
In these equations a bar denotes states of the ordered material. In
addition, the main contributing state of the zinc-blende crystal is
given in parentheses.  As states (\ref{eq:CB_expansion}) and
(\ref{eq:VB_expansion}) are diagonal with respect to the ordering
potential, this removes the potential matrix elements from the
Hamiltonian (\ref{eq:HGL}), but at the price of introducing new $\kdotp$
interactions. The following four momentum matrix elements appear:
\widetext
\begin{mathletters}
  \label{eq:new-P}
  \begin{eqnarray}
    \label{eq:P1s}
    P^{\perp}_{1} &=&  - i \frac{\hbar}{m} 
    \langle \bar\Gamma_{1 \text{c}}(\Gamma_{1 \text{c}})  | p_{x} | 
    \bar\Gamma_{3 \text{v}}^{x}(\Gamma_{5 \text{v}}^{x})  \rangle 
    =  - i \frac{\hbar}{m} 
    \langle \bar\Gamma_{1 \text{c}}(\text{L}_{1 \text{c}})  | p_{x} | 
    \bar\Gamma_{3 \text{v}}^{x}(\text{L}_{3 \text{v}}^{x})  \rangle
    = (\alpha_{\text{v}} \alpha_{\text{c}} + \beta_{\text{v}}
    \beta_{\text{c}}) \PG \\
    \label{eq:P2s}
    P^{\perp}_{2} &=&  - i \frac{\hbar}{m} 
    \langle \bar\Gamma_{1 \text{c}}(\Gamma_{1 \text{c}})  | p_{x} | 
    \bar\Gamma_{3 \text{v}}^{x}(\text{L}_{3 \text{v}}^{x})  \rangle 
    =  - i \frac{\hbar}{m} 
    \langle \bar\Gamma_{1 \text{c}}(\text{L}_{1 \text{c}})  | p_{x} | 
    \bar\Gamma_{3 \text{v}}^{x}(\Gamma_{5 \text{v}}^{x})  \rangle
    = (\alpha_{\text{v}} \beta_{\text{c}} - \beta_{\text{v}}
    \alpha_{\text{c}}) \PG \\
    \label{eq:P1p}
    P^{\parallel}_{1} &=&  - i \frac{\hbar}{m}
    \langle \bar\Gamma_{1 \text{c}}(\Gamma_{1 \text{c}})  | p_{z} | 
    \bar\Gamma_{1 \text{v}}(\Gamma_{5 \text{v}}^{z})  \rangle
    = \alpha_{\text{c}} \PG \\
    \label{eq:P2p}
    P^{\parallel}_{2} &=&  - i \frac{\hbar}{m}
    \langle \bar\Gamma_{1 \text{c}}(\text{L}_{1 \text{c}})  | p_{z} | 
    \bar\Gamma_{1 \text{v}}(\Gamma_{5 \text{v}}^{z})  \rangle
    = \beta_{\text{c}} \PG 
    ,
    \end{eqnarray}
\end{mathletters}
where we have already used relation~(\ref{eq:PG-gleich-PL}). The
momentum matrix elements~(\ref{eq:new-P}) define a standard $\kdotp$
problem of the form of Eq.~(\ref{eq:SG-standard}) for ordered
GaInP$_{2}$,
\begin{equation}
  \label{eq:HGbar}
  H^{\bar \Gamma}(\nk) = 
  \left( 
    \begin{array}{cc}
      H^{\bar\Gamma(\Gamma)}(\nk) 
      & H^{\bar\Gamma(\Gamma)\bar\Gamma(\text{L})}(\nk) \\
      {H^{\bar\Gamma(\Gamma)\bar\Gamma(\text{L})}}^{\dagger}(\nk)
      & H^{\bar\Gamma(\text{L})}(\nk)
    \end{array} 
  \right),
\end{equation}
with
\begin{eqnarray*}
  H^{\bar\Gamma(\Gamma)}(\nk) &=&
  \left(
    \begin{array}{cccc}
      E_{\text{c}}^{(1)} + \frac{\hbar^{2}}{2m}\nkb^{2} +
      {\beta_{\text{c}}}^{2} G {k_{z}}^{2} 
      & i P_{1}^{\perp} \nkb_{x} 
      & i P_{1}^{\perp} \nkb_{y} 
      & i P_{1}^{\parallel} \nkb_{z} \\
      -i P_{1}^{\perp} \nkb_{x} 
      & E_{\text{v}}^{(1)} + \frac{\hbar^{2}}{2m}\nkb^{2} 
      & 0 
      & 0 \\
      -i P_{1}^{\perp} \nkb_{y} 
      & 0 
      & E_{\text{v}}^{(1)} + \frac{\hbar^{2}}{2m}\nkb^{2} 
      & 0 \\
      -i P_{1}^{\parallel} \nkb_{z} 
      & 0 
      & 0 
      & \tilde E^{\Gamma z}_{\text{v}} + \frac{\hbar^{2}}{2m}\nkb^{2}
    \end{array}
  \right),\\
  H^{\bar\Gamma(\text{L})}(\nk) &=&
  \left(
    \begin{array}{ccc}
      E_{\text{c}}^{(2)} + \frac{\hbar^{2}}{2m}\nkb^{2} +
      {\alpha_{\text{c}}}^{2} G {\nkb_{z}}^{2} 
      & i P_{1}^{\perp} \nkb_{x} 
      & i P_{1}^{\perp} \nkb_{y}\\
      -i P_{1}^{\perp} \nkb_{x} 
      & E_{\text{v}}^{(2)}+ \frac{\hbar^{2}}{2m}\nkb^{2}
      & 0 \\
      -i P_{1}^{\perp} \nkb_{y} 
      & 0 
      & E_{\text{v}}^{(2)}+ \frac{\hbar^{2}}{2m}\nkb^{2} 
    \end{array}
  \right),
\end{eqnarray*}
and
\begin{displaymath}
  H^{\bar\Gamma(\Gamma)\bar\Gamma(\text{L})}(\nk) =
  \left(
    \begin{array}{ccc}
      0 
      & - i P_{2}^{\perp} \nkb_{x} 
      & - i P_{2}^{\perp} \nkb_{y} \\
      - i P_{2}^{\perp} \nkb_{x} 
      & 0 
      & 0 \\
      - i P_{2}^{\perp} \nkb_{y}
      & 0 
      & 0 \\
      - i P_{2}^{\parallel} \nkb_{z}
      & 0 
      & 0
    \end{array}
  \right).
\end{displaymath}%
\narrowtext%
Without the approximation of Eq.~(\ref{eq:PG-gleich-PL}) the form of
Hamiltonian~(\ref{eq:HGbar}) would correspond to the general case of a
crystal with $C_{3v}$ symmetry.  The momentum matrix elements
$P^{\perp}_{1}$ and $P^{\perp}_{2}$ determine the effective masses of
$\bar\Gamma_{1 \text{c}}(\Gamma_{1 \text{c}})$ and $\bar\Gamma_{1
  \text{c}}(\text{L}_{1 \text{c}})$ perpendicular to the ordering
direction. The momentum matrix elements $P^{\parallel}_{1}$ and
$P^{\parallel}_{2}$ together with $G$ determine the effective masses
parallel to the ordering direction.  A schematic picture for the
interactions described by the momentum matrix elements is shown in
Fig.~\ref{fig:P-und-P}.


\section{Results and Discussion}
\label{sec:results}

Having set up our model, we can first calculate $|V_{15}|$ and
$|V_{35}|$ for different values of $V_{11}$, and then derive the new
band-edge energies and the expansion coefficients in
Eqs.~(\ref{eq:CB_expansion}) and~(\ref{eq:VB_expansion}). The expansion
coefficients determine the momentum matrix elements (\ref{eq:new-P}),
which, together with the new band-edge energies, yield the effective
masses. The results for the momentum matrix elements and effective
masses are plotted for a range of $|V_{11}|$ up to 0.35~eV. This value
results in a band gap reduction $\Delta E_{\text{BGR}}$ of about
430~meV, which is the theoretical value for the perfectly ordered CuPt
structure.\cite{wezu:98}

Up to now we have not considered the different possibilities for the
relative signs of the potential matrix elements. The sign of $V_{15}$
does not matter since only $|V_{15}|^{2}$ enters into a second order
perturbation theory correction.  Hence only the relative sign of
$V_{11}$ and $V_{35}$, that is $\sigma = \signum(V_{11}/V_{35})$, has to
be determined. We will show that this can be done by appropriate
comparison with experimental results.

\begin{figure}[h]
\epsfxsize=0.45\textwidth \epsfbox{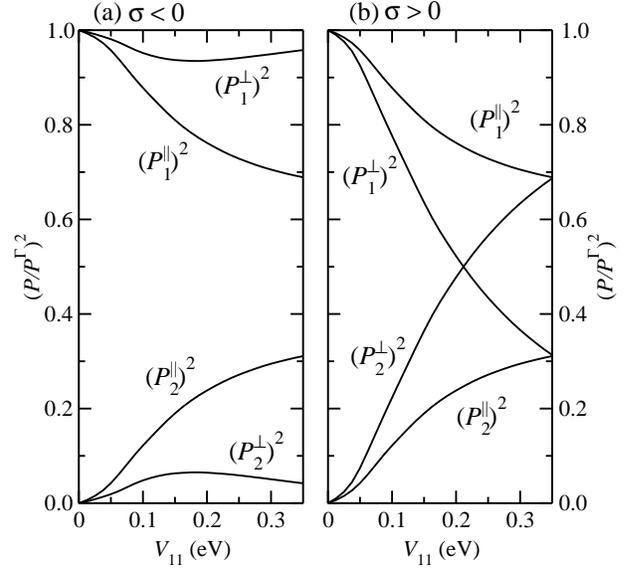}
  \caption{Square of momentum matrix elements [Eq.~(\ref{eq:new-P})] for (a)
    $\sigma = \signum(V_{11}/V_{35}) < 0$ and (b) $\sigma > 0$.}
  \label{fig:P}
\end{figure}

The results for the squares of the four momentum matrix elements
(\ref{eq:new-P}) are shown in Fig.~\ref{fig:P}(a) for $\sigma<0$ and in
Fig.~\ref{fig:P}(b) for $\sigma>0$. The intensity of the optical
transition $\bar\Gamma_{3\text{v}}(\Gamma_{5\text{v}}) \rightarrow
\bar\Gamma_{1\text{c}}(\Gamma_{1\text{c}})$ is proportional to
$|P^{\perp}_{1}|^{2}$, whereas the intensity of the ordering-induced
transition $\bar\Gamma_{3\text{v}}(\Gamma_{5\text{v}}) \rightarrow
\bar\Gamma_{1\text{c}}(\text{L}_{1\text{c}})$ is proportional to
$|P^{\perp}_{2}|^{2}$. Experimental results indicate that the latter
transition is much weaker than the former, even for highly ordered
samples.\cite{kksk:99} Therefore, we can rule out the option $\sigma>0$,
as this would result in approximately the same intensity for these two
transitions.

The difference between $P^{\perp}_{1}$ and $P^{\parallel}_{1}$, and
between $P^{\perp}_{2}$ and $P^{\parallel}_{2}$ should influence the
optical anisotropy of ordered GaInP$_{2}$. This effect has been
neglected in previous calculations.\cite{wezu:94,wezu:94-2} The mirror
symmetry in Figs.~\ref{fig:P}(a) and~\ref{fig:P}(b) with respect to a horizontal
line at $(P/P^{\Gamma})^{2} = 0.5$ is due to the normalization of the
zone center states~(\ref{eq:CB_expansion}) and~(\ref{eq:VB_expansion}).
The matrix elements $(P^{\parallel}_{1})^{2}$ and
$(P^{\parallel}_{2})^{2}$ do not depend on $\sigma$, since the $
\bar\Gamma_{1 \text{v}}(\Gamma_{5 \text{v}}^{z}) $ state is not a
mixture of two different zinc-blende states, and hence there are no
``interference'' terms in Eqs.~(\ref{eq:P1p}) and~(\ref{eq:P2p}). In
Fig.~\ref{fig:P}(b) the relation $(P^{\perp}_{2})^{2} \approx
(P^{\parallel}_{1})^{2}$ and $(P^{\perp}_{1})^{2} \approx
(P^{\parallel}_{2})^{2}$ for $V_{11} = 0.35$~eV are purely accidental.

\begin{figure}[h]
\epsfxsize=0.45\textwidth \epsfbox{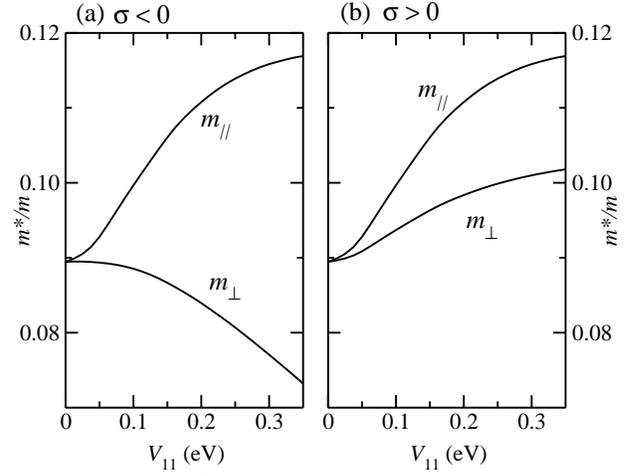}
  \caption{Effective masses of the lowest conduction-band state
    $\bar\Gamma_{1\text{c}}(\Gamma_{1\text{c}})$ for (a) $\sigma =
    \signum(V_{11}/V_{35}) < 0$ and (b) $\sigma > 0$.} 
  \label{fig:masses-G}
\end{figure}

Figure~\ref{fig:masses-G}(a) shows the effective masses of the lowest
conduction-band state $\bar\Gamma_{1\text{c}}(\Gamma_{1\text{c}})$ state
for $\sigma<0$. The effective mass parallel to the ordering direction
$m_{\parallel}$ increases with ordering, whereas the effective mass
perpendicular to the ordering direction $m_{\perp}$ decreases. Within
our model the anisotropy of the effective masses is
$(m_{\parallel}-m_{\perp})/m^{\Gamma}=0.489$ for $|V_{11}|=0.35$~eV,
i.e., for perfect ordering. This value is in good agreement with the
results of Ref.~\onlinecite{fwz:95}. The general trend of the increase
in $m_{\parallel}$ and reduction of $m_{\perp}$ agrees with both
theoretical\cite{fwz:95} and experimental\cite{ezdm:97} results. For
completeness Fig.~\ref{fig:masses-G}(b) shows the effective masses for
$\sigma>0$. It illustrates how important it is to determine $\sigma$
correctly.

\begin{figure}[h]
\epsfxsize=0.45\textwidth \epsfbox{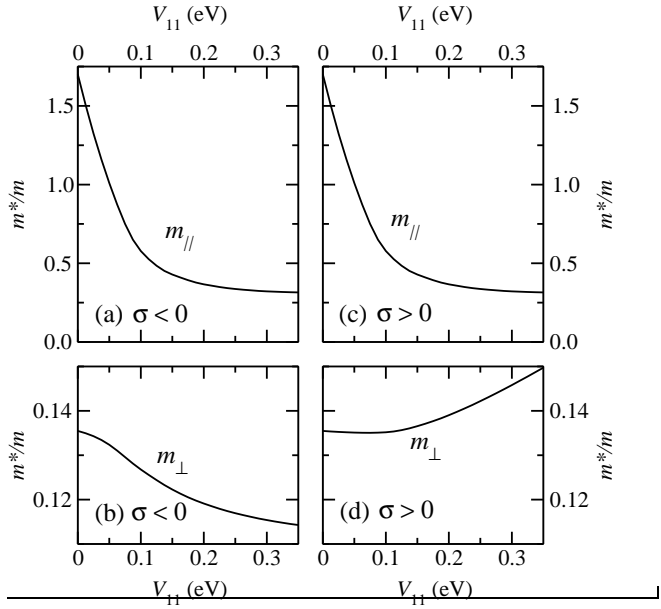}
  \caption{Effective masses of the second-lowest conduction-band state
    $\bar\Gamma_{1\text{c}}(\text{L}_{1\text{c}})$ for (a) and (b) $\sigma =
    \signum(V_{11}/V_{35}) < 0$, and (c) and (d) $\sigma > 0$.} 
  \label{fig:masses-L}
\end{figure}

The predictions of our model for the effective masses of the
$\bar\Gamma_{1\text{c}}(\text{L}_{1\text{c}})$ state are shown in
Figs.~\ref{fig:masses-L}(a) and~\ref{fig:masses-L}(b), again for $\sigma<0$. The most striking
feature is the decrease in the effective mass parallel to the ordering
direction from 1.7 to less than 0.4. The effective mass perpendicular to
the ordering direction shows an increase, comparable in magnitude to the
changes for the $\bar\Gamma_{1\text{c}}(\Gamma_{1\text{c}})$ effective
masses.  To the best of our knowledge, the present work is the first
investigation of the effective masses of this second lowest conduction
band in the ordered material. For completeness
Figs.~\ref{fig:masses-L}(c) and~\ref{fig:masses-L}(d) show the effective masses for $\sigma>0$.
Note that $m_{\parallel}$ does not depend on $\sigma$ in
Figs.~\ref{fig:masses-G} and~\ref{fig:masses-L}. This can be easily
understood, as these masses are determined by the terms proportional to
$G$ in Eq.~(\ref{eq:HGbar}) and by $(P^{\parallel}_{1})^{2}$ or
$(P^{\parallel}_{2})^{2}$, respectively, which are independent of
$\sigma$.

\section{Summary and Outlook}
\label{sec:summary}

In conclusion, we have presented a general formalism which extends
standard $\kdotp$ theory to periodic perturbations, which are rapidly
oscillating on a length scale of a few lattice constants.  We choose a
suitable complete and orthonormal basis that makes it possible to
consider explicitly the interactions due to the perturbation. Our ansatz
can be readily combined with other extensions of $\kdotp$ theory such as
for the inclusion of strain and spin-orbit interaction, thereby
retaining the simple analytic formulas of $\kdotp$ theory.  When the
period of the perturbation increases, more $\set$ points in the
Brillouin zone have to be considered in our ansatz, increasing the
number of potential matrix elements $\VnnKK$.  However, the number of
independent parameters can be significantly reduced using symmetry
arguments, as illustrated in our discussion of ordered GaInP$_{2}$.  If
it is desirable to decrease the number of parameters further,
one can in a perturbative way restrict the calculation to a
subset $\{\set\}$ containing only extremal points of the energy
dispersion, which usually are most important.  Alternatively, one can
calculate the potential matrix elements according to their microscopic
definition [Eq.~(\ref{eq:VnnKK})] using wave fuctions from, e.g., a
pseudopotential calculation for the unperturbed system.

As an example,
we calculate the effective masses in the lowest two conduction bands of
spontaneously ordered GaInP$_{2}$ as a function of the degree of
ordering. For the lowest conduction band we find qualitatively good
agreement between our results, first-principle calculations and
experimental data. We also find the momentum matrix element between
conduction- and valence-band states to be anisotropic, which influences
the optical anisotropy of ordered GaInP$_{2}$.  Although we have
calculated the curvatures of the conduction bands only, our approach can
also be applied to the valence band. To do this, a consistent set of band
parameters is required and spin-orbit interaction should be taken into
account.  We expect, e.g., that the different signs of the 
curvature of the valence band parallel to the ordering direction at
$\Gamma$ and L points cause an increase in the heavy-hole mass parallel
to the ordering direction.  Besides GaInP$_{2}$, which we have treated
here, natural short-period superlattices occur in many different
semiconductor alloys (cf., e.g., Ref \onlinecite{rats:94}), and our
method is well suited to describe these systems.


\begin{thebibliography}{10}
\bibitem{notation}
Note that we have used lower case letters ($\ak$, $\abasis$, $\ac$, etc.) in
  the above discussion to denote quantities in standard $\kdotp$ theory and
  upper case letters ($\nk$, $\nbasis$, $\nc$, etc.) to denote the
  corresponding quantity in our generalized theory.
\bibitem{theta}
Note that the value of $\theta$ in Ref.~\onlinecite{kksk:99} differs from the
  value here, though this has only a marginal effect for our application.
\end{thebibliography}

\input{epsf}

\begin{table}[h]
  \caption{Energies, effective masses, and momentum matrix elements or
    remote bands contribution, respectively, for $\Gamma_{1\text{c}}$,
    $\text{L}_{1\text{c}}$, and $\text{L}_{3\text{v}}$. The energy zero is
    the valence-band maximum $\Gamma_{5\text{v}}$.}
  \label{tab:E-m-P}
  \begin{tabular}{clll}
    State &
    \multicolumn{1}{c}{Energy} &
    \multicolumn{1}{c}{Effective mass} &
    \multicolumn{1}{c}{Matrix element} \\ 
     &
    \multicolumn{1}{c}{(eV)} &
    \multicolumn{1}{c}{($m$)} &
     \\
    \hline
    $\Gamma_{1\text{c}}$ & 
    $E^{\Gamma}_{\text{c}} = 2$.024  & 
    $m^{\Gamma} = 0$.0899  &
    $\PG = 8$.83~eV\AA \\
    $\text{L}_{1\text{c}}$ & 
    $E^{\text{L}}_{\text{c}} = 2$.250  & 
    $m^{\text{L}}_{\perp} = 0$.1349  &
    $\PL = 8$.88~eV\AA \\
     & 
     &
    $m^{\text{L}}_{\parallel} = 1$.699  &
    $G = -1$.57~eV\AA$^{2}$ \\
    $\text{L}_{3\text{v}}$ & 
    $E^{\text{L}}_{\text{v}} = -0$.978  & 
    &
  \end{tabular}
\end{table}

\widetext

\end{document}